\documentclass[11pt,en,bibtex]{elegantpaper}

\usepackage{subcaption}
\usepackage{abstract}
\usepackage{comment}
\usepackage{wrapfig}
\usepackage{float}
\usepackage{hyperref}
\usepackage{xcolor}
\usepackage{tcolorbox}
\usepackage{authblk}
\usepackage{lineno}  
\title{A Test System for the JUNO 20-inch PMTs Prior to Installation}

\author[1,2]{Zhaoyuan Peng}
\author[1,2]{Haojie Dong}
\author[1]{Kaile Wen}
\author[3]{Xinzhou Guo}
\author[1]{Yanfeng Li}
\author[1]{Songyi Li}   
\author[3]{Zeyuan Feng}
\author[1]{Wan Xie}
\author[1]{Shenghui Liu}
\author[1,2]{Chao Chen}
\author[1,2]{Xiaochuan Xie}
\author[1]{Jun Hu}
\author[1]{Lei Fan}
\author[1]{Zhonghua Qin}

\affil[1]{Institute of High Energy Physics, Beijing 100049, China}
\affil[2]{University of Chinese Academy of Sciences, Beijing 100049, China}
\affil[3]{Southern University of Science and Technology, Shenzhen, China}

\usepackage{array}

\begin{document}

\maketitle

\begin{abstract}
The JUNO experiment requires an excellent energy resolution of 3\% at 1 MeV. To achieve this objective, a total of 20,012 20-inch photomultiplier tubes (PMTs) will be deployed for JUNO, comprising 15,012 multi-channel plate (MCP) PMTs and 5,000 dynode PMTs. Currently, JUNO is in the process of detector installation, with PMTs being installed from the top to the bottom of the stainless-steel structure located in the underground experimental hall. In order to validate the functionality of the PMTs and ensure there are no malfunctions prior to installation, a test system has been established at the JUNO site, and testing is being conducted. This paper presents an overview of the test system and reports on the initial test results.
\keywords{JUNO, Potted PMT, PMT Test}
\end{abstract}

\section{Introduction and Motivation}
\begin{figure}[htb]
\centering
\includegraphics[width=0.9\textwidth]{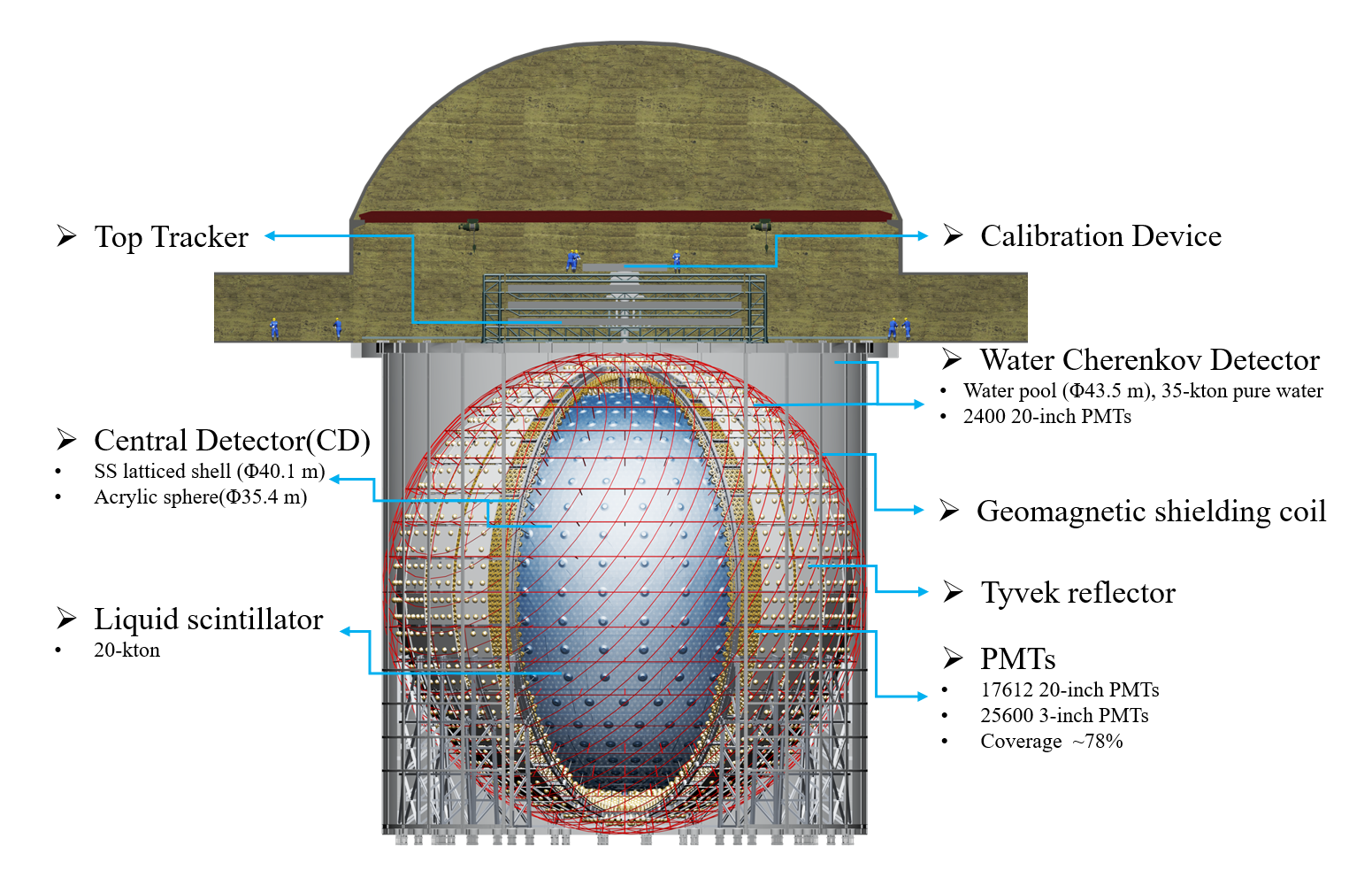}
\caption{Schematic view of the JUNO detector}
\label{fig:JUNO Detector}
\end{figure}

The Jiangmen Underground Neutrino Observatory (JUNO) \cite{1} is a multipurpose neutrino experiment, with its physics motivations including the determination of neutrino mass ordering, precise measurement of oscillation parameters, and the study of atmospheric neutrinos, solar neutrinos, geoneutrinos, and supernova neutrinos, etc \cite{2}. 
The JUNO detector, as depicted in Fig. \ref{fig:JUNO Detector}, is a large liquid scintillator detector comprised of several subsystems, including a central detector \cite{3}, a water Cherenkov detector \cite{4}, and a top tracker \cite{5}. 
The central detector features a spherical acrylic vessel with a diameter of 35.4 m, supported by a stainless-steel latticed shell with a diameter of 40.1 m, which contains 20 ktons of liquid scintillator \cite{6}. Within the central detector, there are 17,612 20-inch PMTs and 25,600 3-inch PMTs,  served as photosensors \cite{CAO2021165347,Abusleme2022}. The water Cherenkov detector is a cylindrical water pool with 43.5 m in diameter and 44 m in depth, equipped with 2,400 20-inch PMTs, and filled with 35 ktons of ultra-pure water for cosmic muon vetoing and environmental radioactivity shielding. Additionally, the top tracker consists of three layers of plastic scintillators designed to precisely tag muon tracks \cite{inproceedings}. A calibration system is also necessary to calibrate the detector before and during JUNO's operation \cite{Abusleme2021}. 

Among the 20,012 20-inch PMTs, 15,012 are MCP PMTs produced by NNVT \cite{12}, while the remaining 5,000 are dynode PMTs manufactured by Hamamatsu \cite{13}. 
To evaluate the performance and characterize the PMTs, a facility was established approximately 150 km from the JUNO site, namely the Pan-Asia PMT test and potting station (hereafter referred as Pan-Asia) \cite{Abusleme2022}. 
An acceptance test was conducted for each PMT upon its arrival at Pan-Asia. 
This test involved measuring a comprehensive list of parameters, including photon detection efficiency (PDE), dark count rate (DCR), single photoelectron charge spectra (SPE), gain, transit time spread (TTS), and after-pulse, etc. 
Any PMTs that failed to meet the quality standards were returned to the manufacturers. 
This test campaign commenced in 2017 and continued for four years until the arrival of all PMTs. Details regarding the test facility and the test results can be found in the Ref. \cite{8,9}.

After the completion of the acceptance tests, the qualified PMTs were approved for the JUNO, and a waterproof potting process for these PMTs commenced. 
The potting of all PMTs took more than two years, from July 2019 to August 2021, and sampling tested \cite{14}. Subsequently, the PMTs were integrated with protective covers to prevent chain implosion when submerged in water \cite{15,16}. Finally, the installation of the PMTs began in October 2022, and the PMTs were transported in batches from the Pan-Asia to the JUNO site.

\begin{figure}[htb]
\centering
\includegraphics[width=0.6\textwidth]{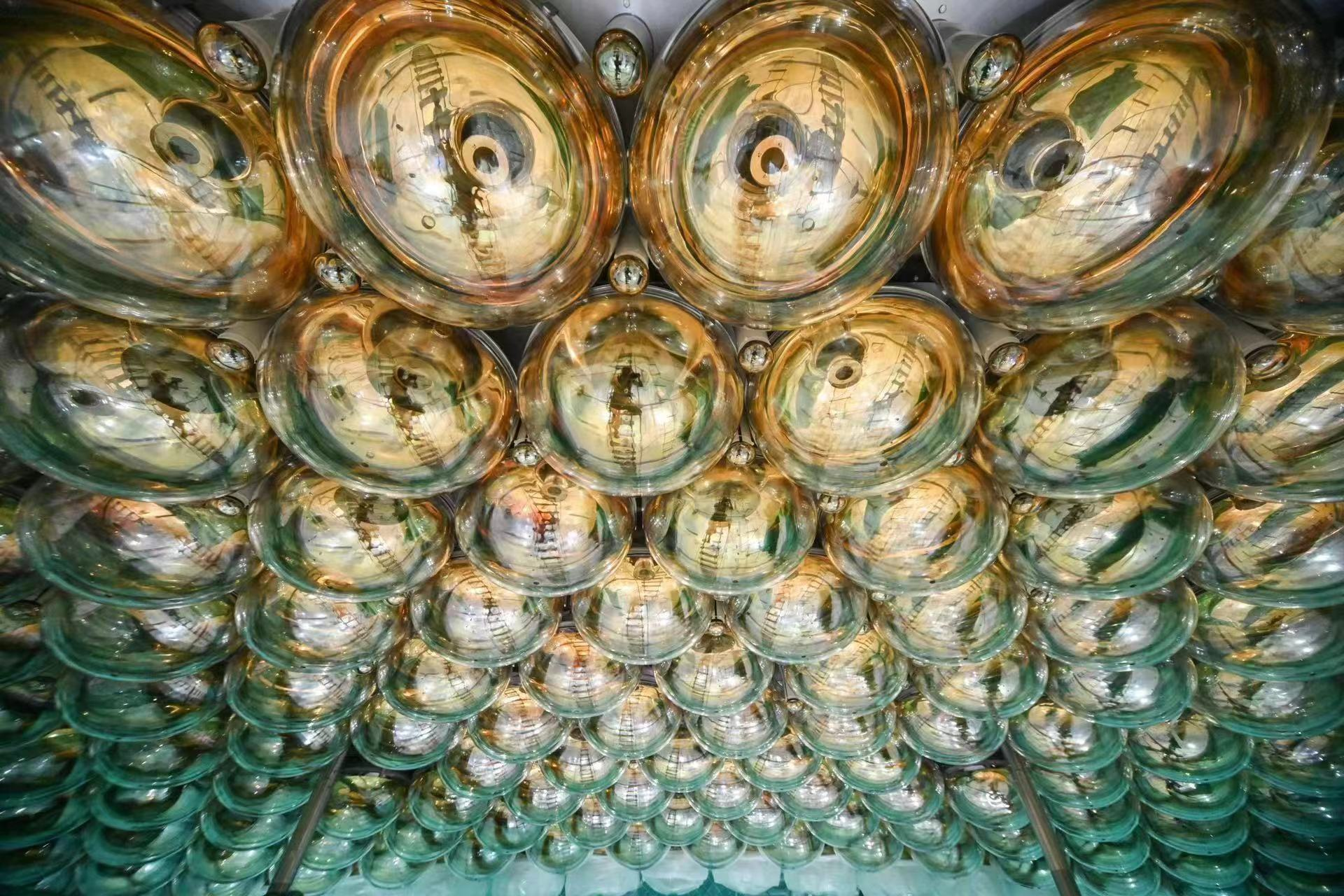}
\caption{A section of the installed PMTs}
\label{fig:PMT layout}
\end{figure}

The optical coverage requirement for the PMT installation in JUNO is set at 78\%, which necessitates maintaining only a few millimeters of gap between adjacent PMTs, as illustrated in Fig. \ref{fig:PMT layout}, where a section of the installed PMTs is shown. Due to this compact layout, it is nearly impossible to replace a PMT once it has been installed, even though a malfunction is detected afterwards. Moreover, since the PMTs undergo several operations after the initial acceptance test, a final functionality validation is essential before they are mounted onto the stainless-steel latticed shell. To this end, a specialized test system was designed and constructed in the surface assembly building (SAB) of JUNO, which enables the mass test of the PMTs prior to their installation at the underground. 

\section{Description of the Test System}
\begin{figure}[htp]
\centering
\includegraphics[width=0.8\textwidth]{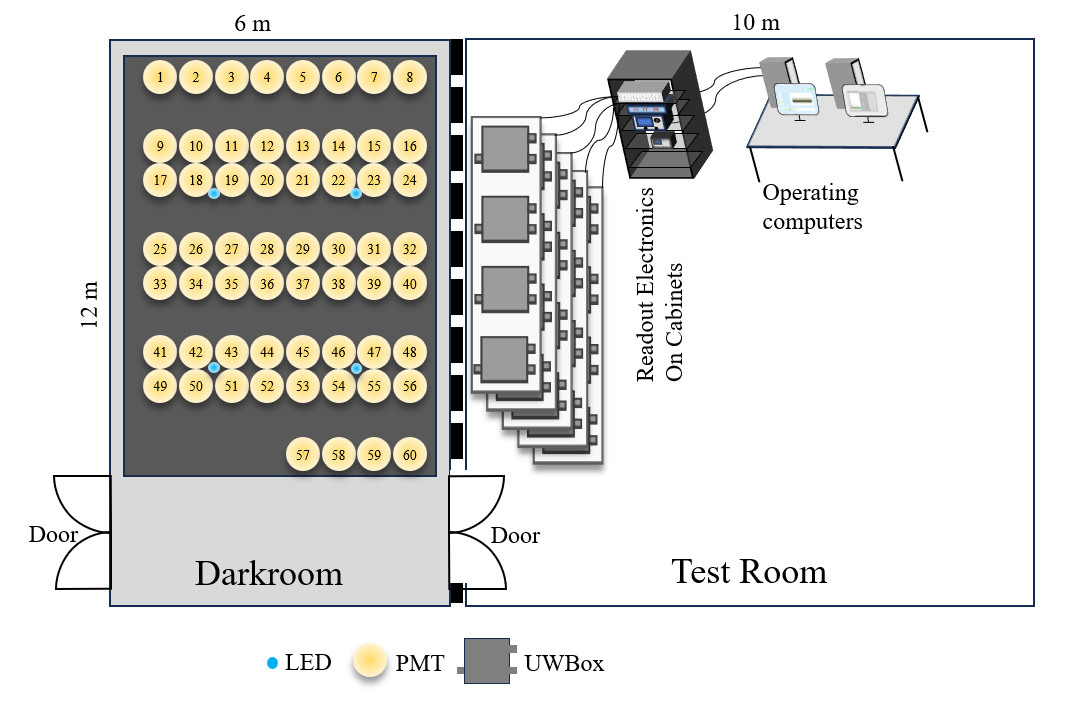}
\caption{The layout of the test system}
\label{fig:Testroom layout}
\end{figure}

As previously mentioned, the test system in JUNO SAB is designed to verify no malfunction of PMTs before installation, therefore it is unnecessary to perform a comprehensive re-evaluation of all parameters tested at Pan-Asia. 
Instead, only the DCR and the charge spectra from LEDs were selected for the validation tests, as they indicate the PMT's noise level and its behavior in the presence of light, respectively. 
Moreover, to ensure that the PMT test results closely reflect the final performance in JUNO and provide valuable references for future operations, the testing conditions must closely mimic the actual JUNO environment. 
This includes factors such as mitigation of the geomagnetic field, utilization of the final JUNO readout electronics, and a similar data acquisition system.
Fig. \ref{fig:Testroom layout} illustrates the overall layout of the test system, which is divided into two rooms: the dark room, where the PMTs are housed, and the test room, which contains the readout electronics and the test platform. 
A total of 60 PMTs can be accommodated in the dark room for parallel testing. 
Each component of the test system will be elaborated in the subsequent sections. 

\subsection{The Dark Room and the Geomagnetic Field Shielding}
\begin{figure}[htb]
\captionsetup[subfigure]{labelformat=parens}
\centering
\subcaptionbox{The empty dark room\label{fig:the empty darkroom}}{
    \includegraphics[width=0.45\linewidth]{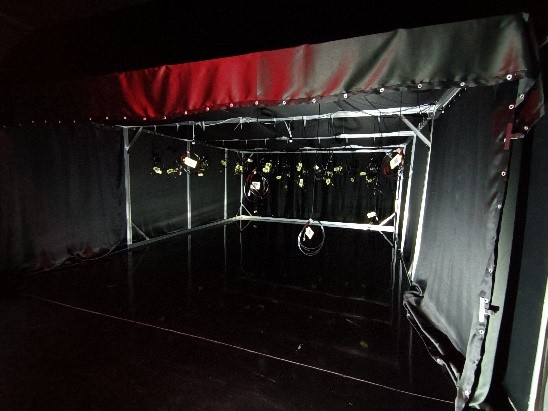}
}
\hspace{0.05\linewidth} 
\subcaptionbox{PMTs in the dark room\label{fig:PMTs in the darkroom}}{
    \includegraphics[width=0.45\linewidth]{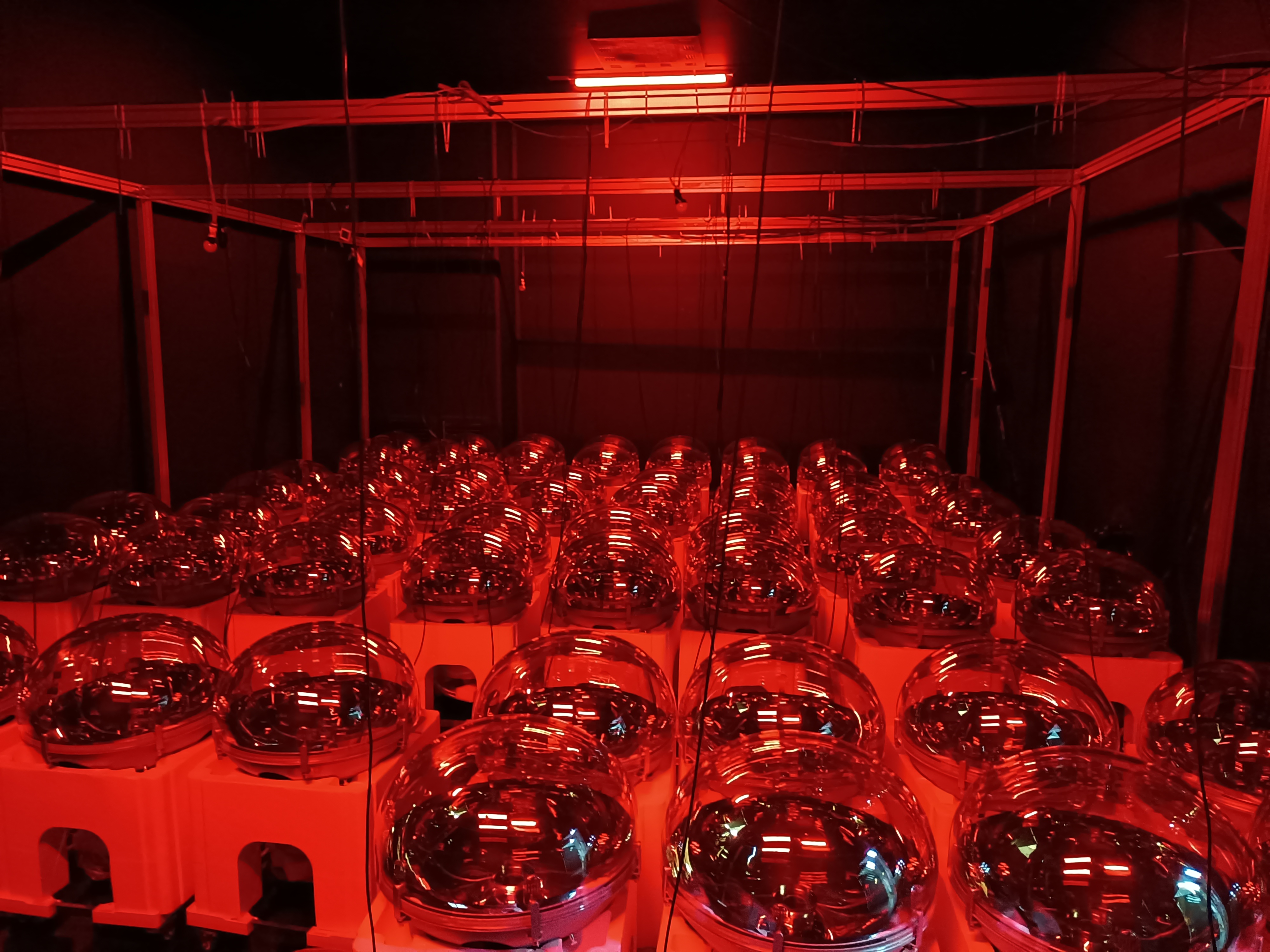}
}
\caption{The dark room and PMTs to be tested}
\label{fig:SAB test darkroom and PMTs}
\end{figure}

A dark room of 12 m by 6 m was constructed for this test. The room was meticulously designed to prevent light leakage from the outside, given its relatively expansive area. The side and top walls of the room were lined with a layer of black wallpaper, and the doors were fitted with black curtains on both the inside and outside. Additionally, an inner region in the room was created using an aluminum frame covered with curtains, which is able to further minimize light leakage, thereby satisfying the testing requirements. To match installation speed, the dark room was designed to accommodate 60 PMTs, with two tests conducted each day, allowing for a total of 120 PMTs to be tested daily. The images presented in Fig. \ref{fig:SAB test darkroom and PMTs} depict the empty dark room and the PMTs scheduled for testing in the room.

The geomagnetic field significantly affects the performance of the PMTs, particularly in terms of PDE, gain, and TTS, so a design with shielding coils (as shown in Fig. \ref{fig:JUNO Detector}) was implemented to the JUNO detector to reduce the geomagnetic field by 90\%.
To minimize the geomagnetic field in the dark room and test the PMTs under the same conditions as in the detector, various methods were considered and tested in the dark room to shield the geomagnetic field. 
These methods included a full shielding of the inner region, individual shielding for each PMT, the use of single or double layers of shielding, and the application of different shielding materials.

\begin{figure}[htp]
\centering
\begin{subfigure}[b]{0.3\textwidth}
    \includegraphics[width=\textwidth,height=5cm]{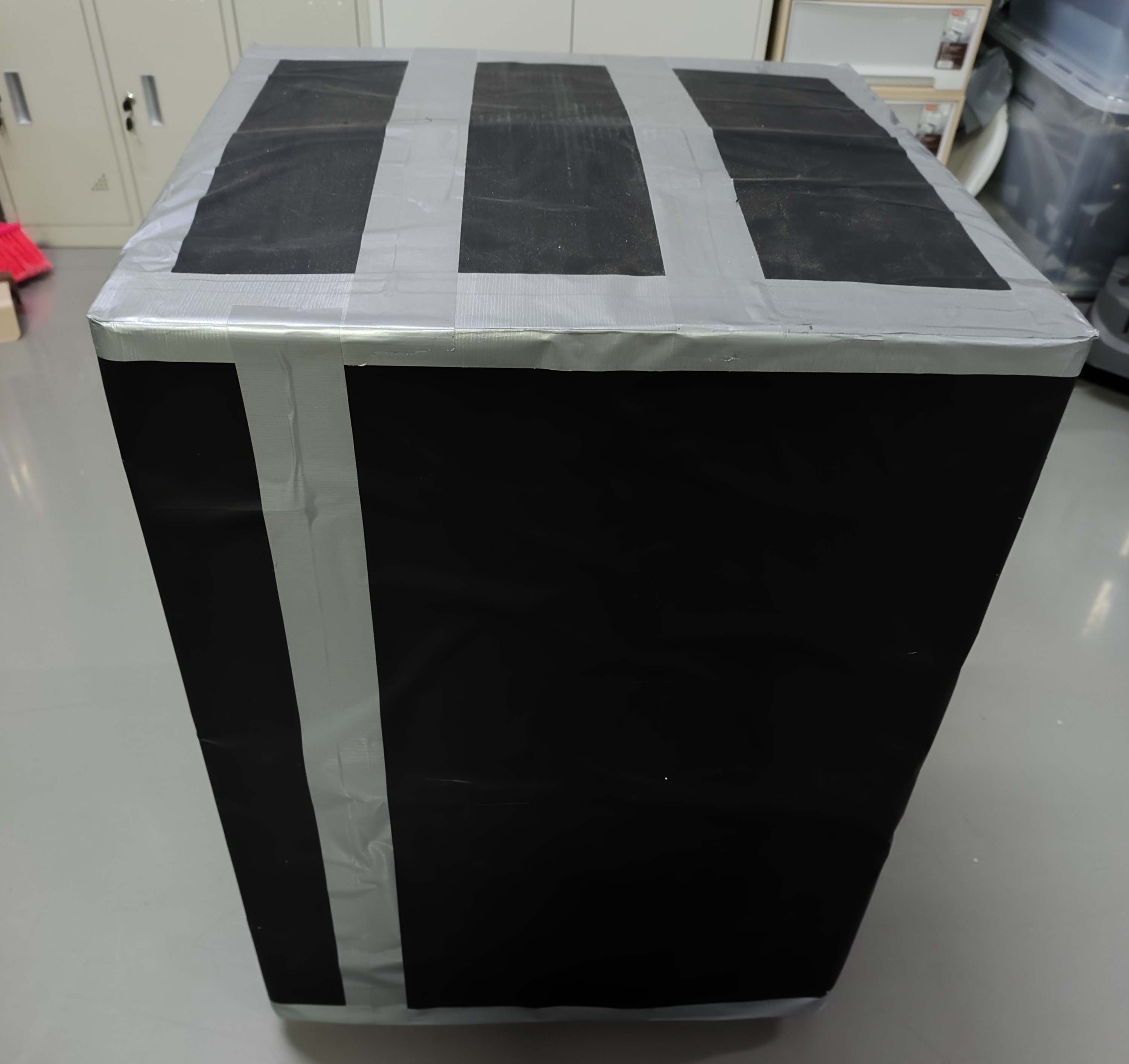}
    \caption{}
    \label{fig:shieldingbox1}
\end{subfigure}
\hfill
\begin{subfigure}[b]{0.3\textwidth}
    \includegraphics[width=\textwidth,height=5cm]{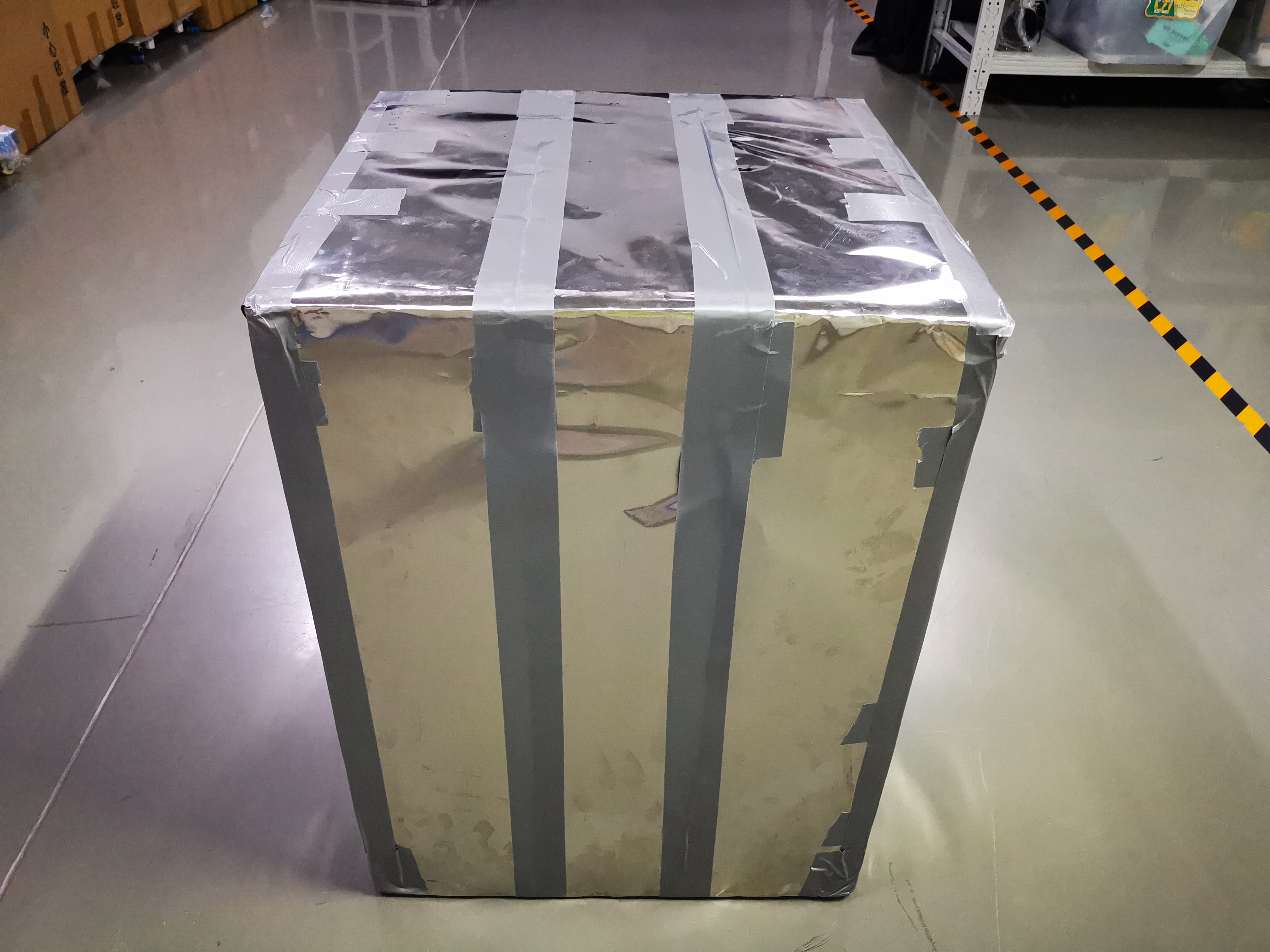}
    \caption{}
    \label{fig:shieldingbox2}
\end{subfigure}
\hfill
\begin{subfigure}[b]{0.3\textwidth}
    \includegraphics[width=\textwidth,height=5cm]{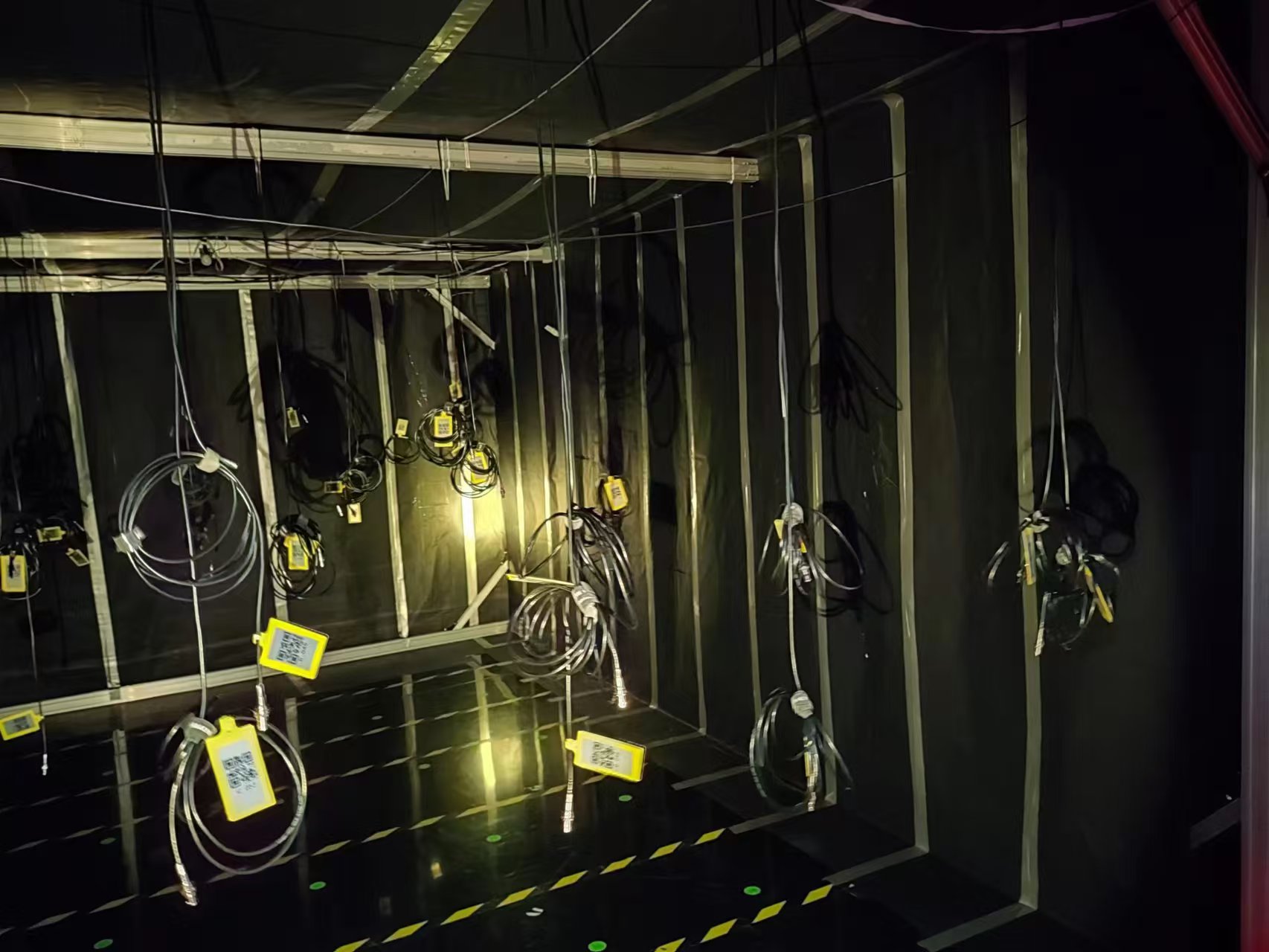}
    \caption{}
    \label{fig:shielding3}
\end{subfigure}
\caption{(a):1-layer nanocrystals pasted on PMT cardboard box. (b):1-layer permalloy pasted on PMT cardboard box. (c):1-layer nanocrystals applied to inner darkroom.}
\label{fig:shieldings}
\end{figure}

\begin{figure}[htb]
\centering
\includegraphics[width=0.8\textwidth]{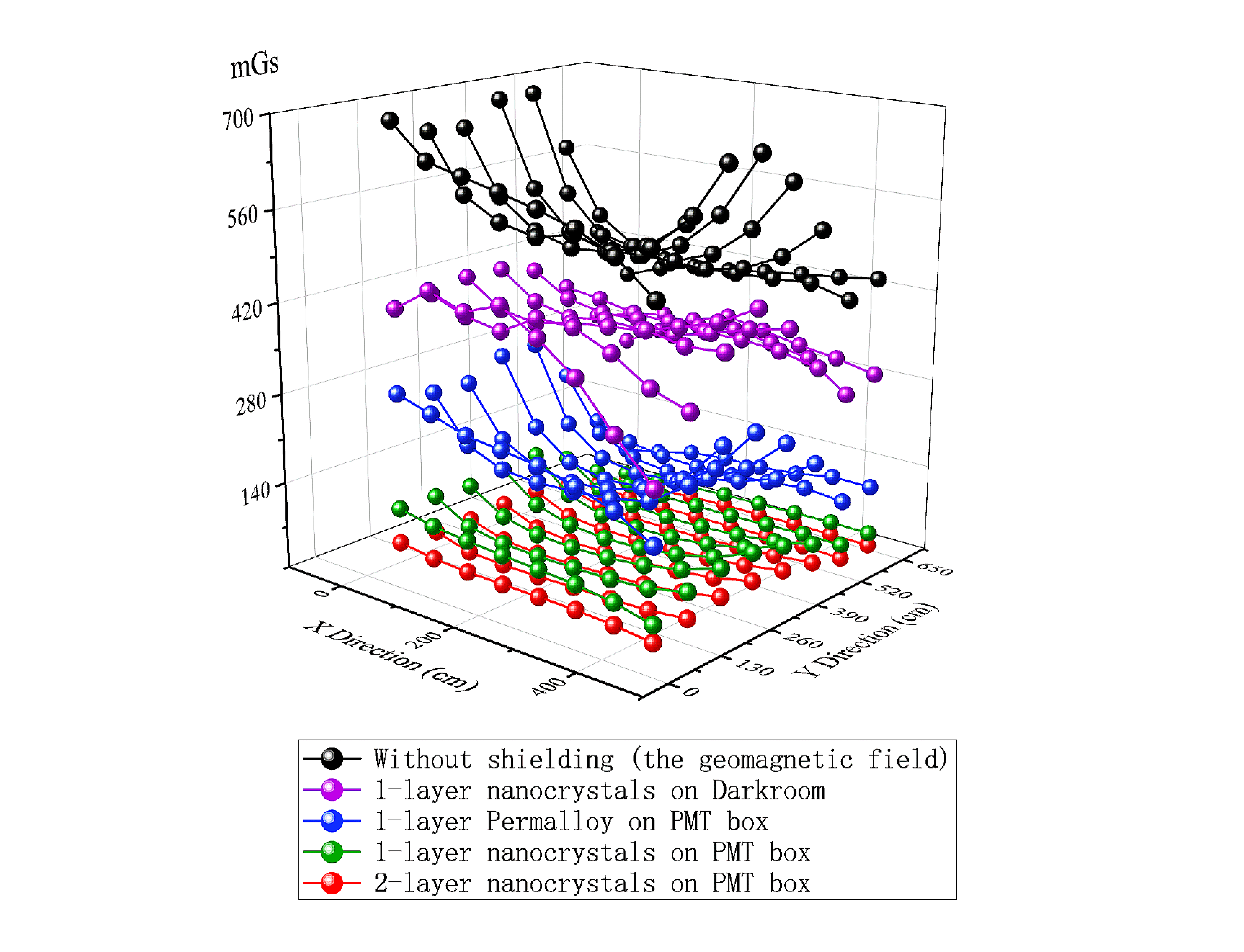}
\caption{The residual geomagnetic field for different shielding method}
\label{fig: residual EMF}
\end{figure}

Fig. \ref{fig:shieldings} shows the shieldings for different methods.
Fig. \ref{fig: residual EMF} presents the detailed distribution of the residual magnetic field in the dark room for the different shielding methods employed. 
The data indicate that values near the side walls of the dark room are generally higher than those measured at the center, likely due to the walls partially being made of metal. 
Average values for the various shielding methods are summarized in Tab. \ref{tab:GeoMag}.

Among these methods, the 2-layer shielding with nanocrystals exhibited the best performance; however, a 1-layer nanocrystals is sufficient for our test as a final magnetic shielding choice, providing approximately 90\% attenuation of the geomagnetic field \cite{nanocrystalline}. 
This 1-layer method also reduces costs associated with the shielding setup. 
To accommodate the shape of the PMTs, the nanocrystals layers were directly applied to the PMT cardboard boxes, as depicted in Fig. \ref{fig:shieldingbox1}. 
In total, 60 individual shielding boxes of 1-Layer Nanocrystals were made for daily test.

\begin{table}[htp]
\centering
\caption{The residual geomagnetic field measurements.}
\label{tab:GeoMag}
\begin{tabular}{c|c} 

Shielding methods & Average value after shielding (mGs) \\
\hline 
Without shielding (the geomagnetic field) & 528  \\ 
1-layer nanocrystals on Darkroom & 374  \\ 
1-layer Permalloy on PMT box & 174 \\ 
1-layer nanocrystals on PMT box & 62 \\ 
2-layer nanocrystals on PMT box & 31 \\ 
\end{tabular}
\end{table}

\subsection{Light Source and Readout Electronics}

\begin{figure}[htb]
\centering
\includegraphics[width=0.8\textwidth]{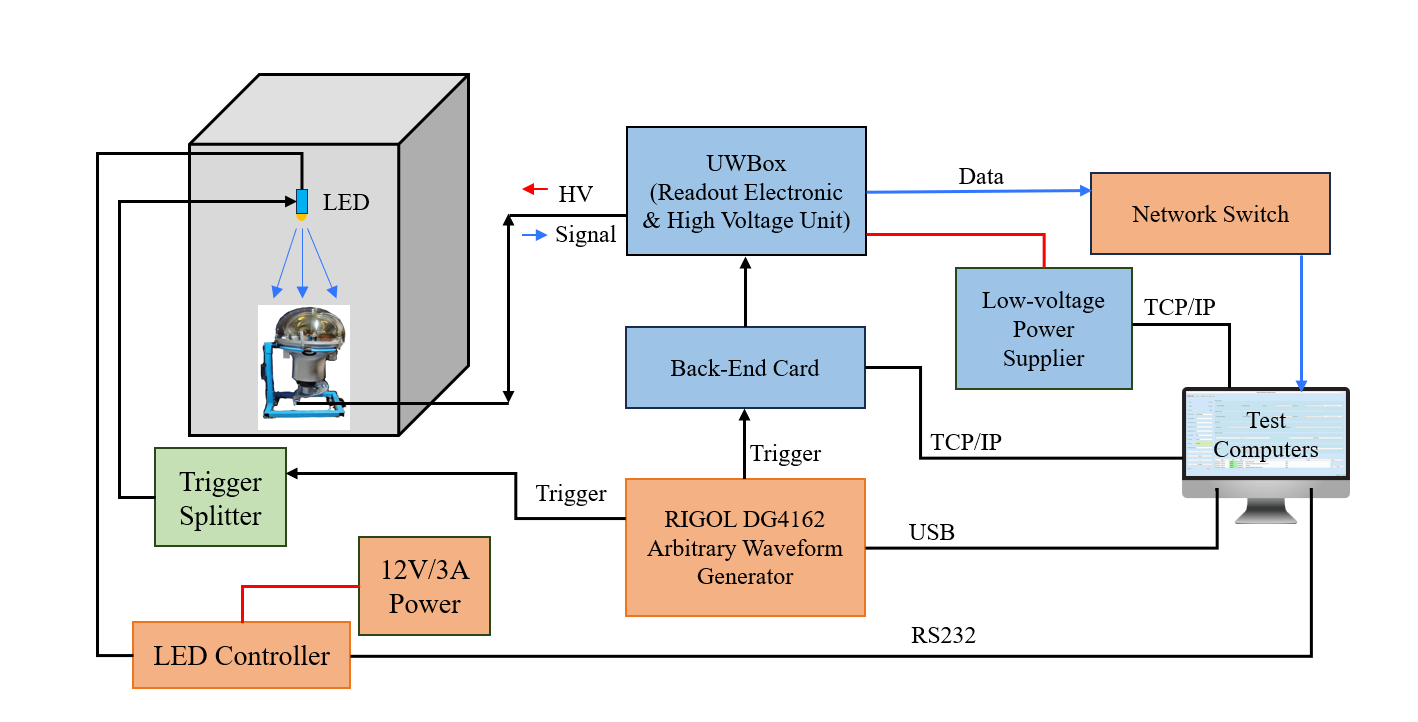}
\caption{Scheme of light source and readout electronics}
\label{fig:test single PMT}
\end{figure}

\begin{figure}[hbt]
\centering
\includegraphics[width=0.5\textwidth]{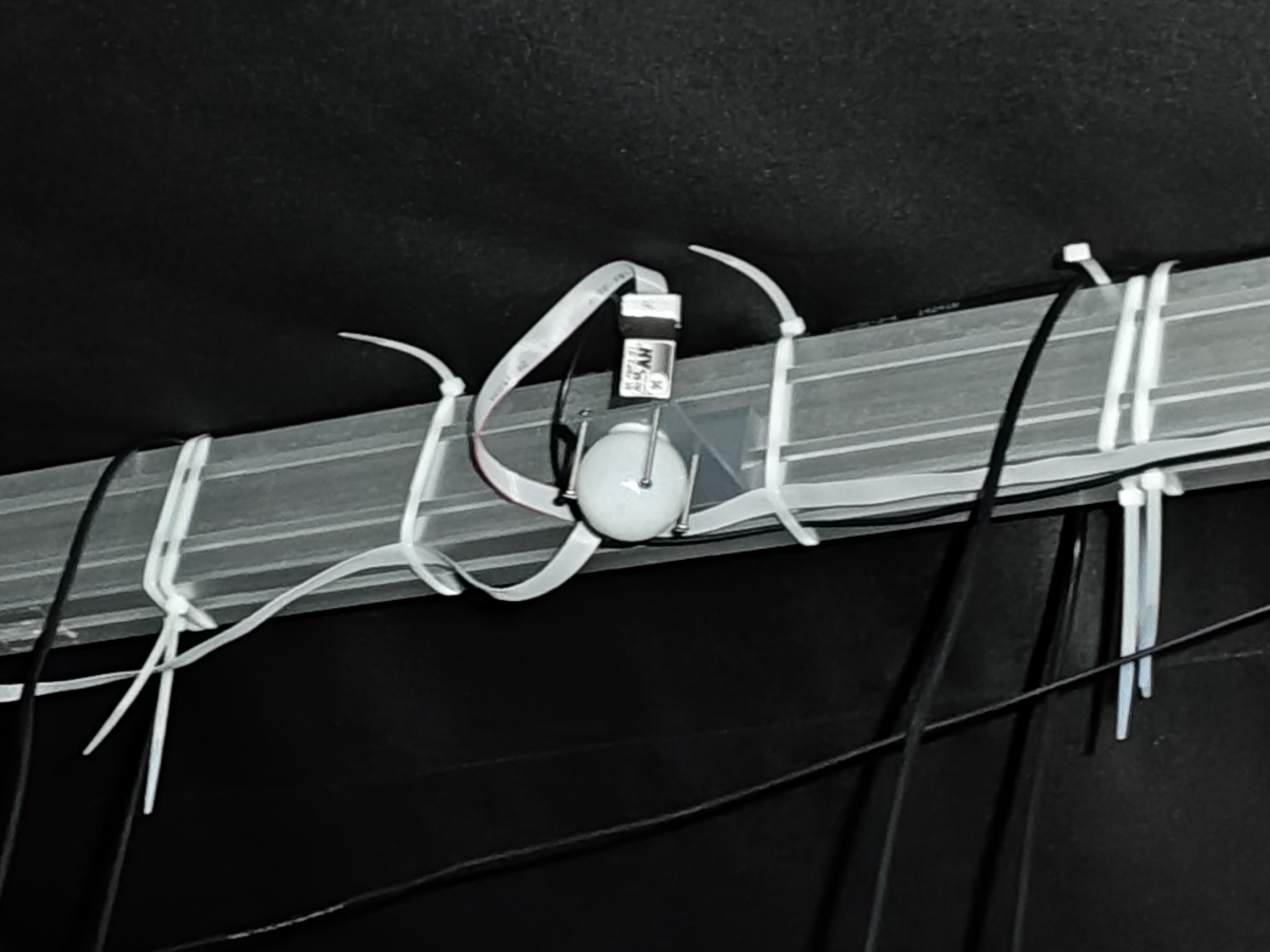}
\caption{A LED with diffused ball positioned at the top of the darkroom.}
\label{fig:LED}
\end{figure}

Fig. \ref{fig:test single PMT} illustrates the schematic overview of the light source and readout electronics of the test system. To assess the PMT's response to light, light-emitting diodes (LEDs) were positioned at the top of the inner region of the dark room. Given the room's considerable size, a single LED could not adequately cover all PMTs. Consequently, four LEDs were distributed throughout the space. One LED is depicted in Fig. \ref{fig:LED}, where the head of the LED is outfitted with a diffuse ball to enhance light distribution.

The LEDs employed in this setup are the same as those used in the container system at Pan-Asia. They feature a self-stabilizing mechanism that maintains constant intensity through a feedback loop in the circuit. Further details about the LED specifications can be found in Ref. \cite{8}. The output intensity of LED can be adjusted from several thousand photons to single photon level. For this test, the intensity was set to emit a few tens of photons. The LEDs operate at a wavelength of approximately 420 nm, coinciding with the wavelength of scintillating light produced in JUNO.

As illustrated in Fig. \ref{fig:test single PMT}, the JUNO readout electronics is utilized in this test to mimic the noise and signal conditions of the actual JUNO setup.  
This choice is also influenced by the fact that the PMTs have been potted and fitted with customized waterproof connectors for their integration with the final electronics.
The JUNO readout electronics comprise the front-end electronics, which are submerged in water, and the back-end electronics located in the experimental hall cabinets. 
The front-end electronics include Analog-to-Digital Converters (ADCs), Global Control Units (GCUs), and High Voltage Units (HVUs), all housed in underwater boxes (UWBoxes). Each UWBox is connected to three PMTs. The back-end electronics primarily consist of the Back-End Card (BEC). A detailed design of the JUNO electronics can be found in Ref. \cite{10,10_2}.

\begin{figure}[htb]
\centering
\includegraphics[width=0.5\textwidth]{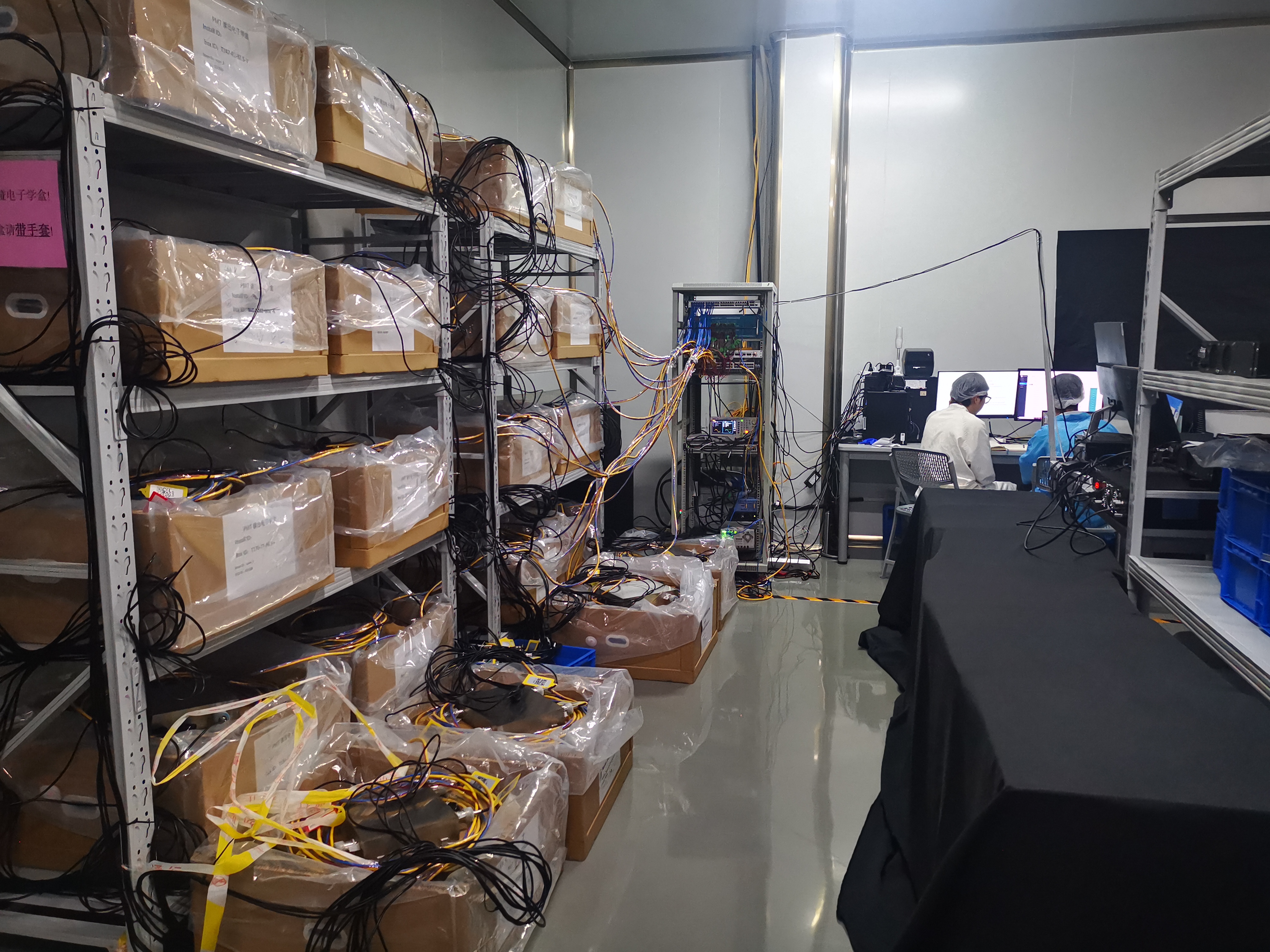}
\caption{Readout electronics and test platform in test room}
\label{fig:readout_elec}
\end{figure}

In this test system, which simultaneously tests 60 PMTs, a total of 20 UWBoxes and one BEC are required, with power supplied by a single low-voltage power supply (LVP).
The UWBoxes were running in air and the room temperature is controlled at 21 °C (the same as JUNO running temperature).
A photograph of the readout electronics and test platform in the test room is provided in Fig. \ref{fig:readout_elec}.

For the measurement of the DCR, an algorithm counts signals over a given threshold developed on the FPGA of the readout electronics and stores the data in a register for subsequent access by the test program. A self-trigger mode is employed for DCR measurement. 

For the charge spectra measurement, an external trigger mode is employed, generated by a RIGOL DG4162 Arbitrary Waveform Generator. 
This device provides the necessary NIM signals for the LEDs and BEC with minimal jitter ($\leq 50 \text{ ps}$) and is remotely controlled via USB interface. Several coaxial signal cables (RG142, with varying lengths and BNC connectors) are used to synchronize and distribute the trigger signals to the BEC and LEDs. Additionally, CAT5E F/UTP Ethernet cables connect the BEC and LVP to the computer via TCP-IP protocols.

\subsection{Data Acquisition and Control System}

\begin{figure}[htb]
\centering
\includegraphics[width=0.95\textwidth]{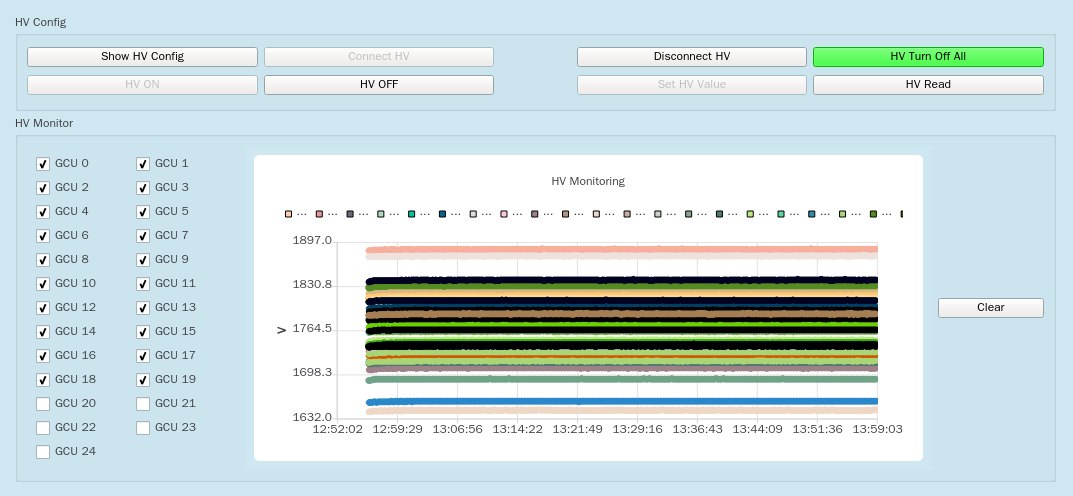}
\caption{High voltage controller and monitor implemented in DAQ}
\label{fig:HV}
\end{figure}

A prototype data acquisition (DAQ) system developed for the JUNO is employed to collect data from the readout electronics. This system acquires the digitized waveform streams from the 20 GCUs in parallel and stores them on a local computer based on TCP-IP protocol. Customized functions, including waveform display, high voltage control and monitoring, are also implemented in this prototype system. As demonstrated in Fig. \ref{fig:HV}, the high voltages for the 60 PMTs (connected to 20 GCUs) were activated, and these values were monitored throughout the test. A detailed introduction to this system can be found in Ref \cite{11}. The hardware component consists of a 24-port network switch that transfers waveform data to the local computer and facilitates the exchange of necessary information.

\begin{figure}[htb]
\centering
\includegraphics[width=0.95\textwidth]{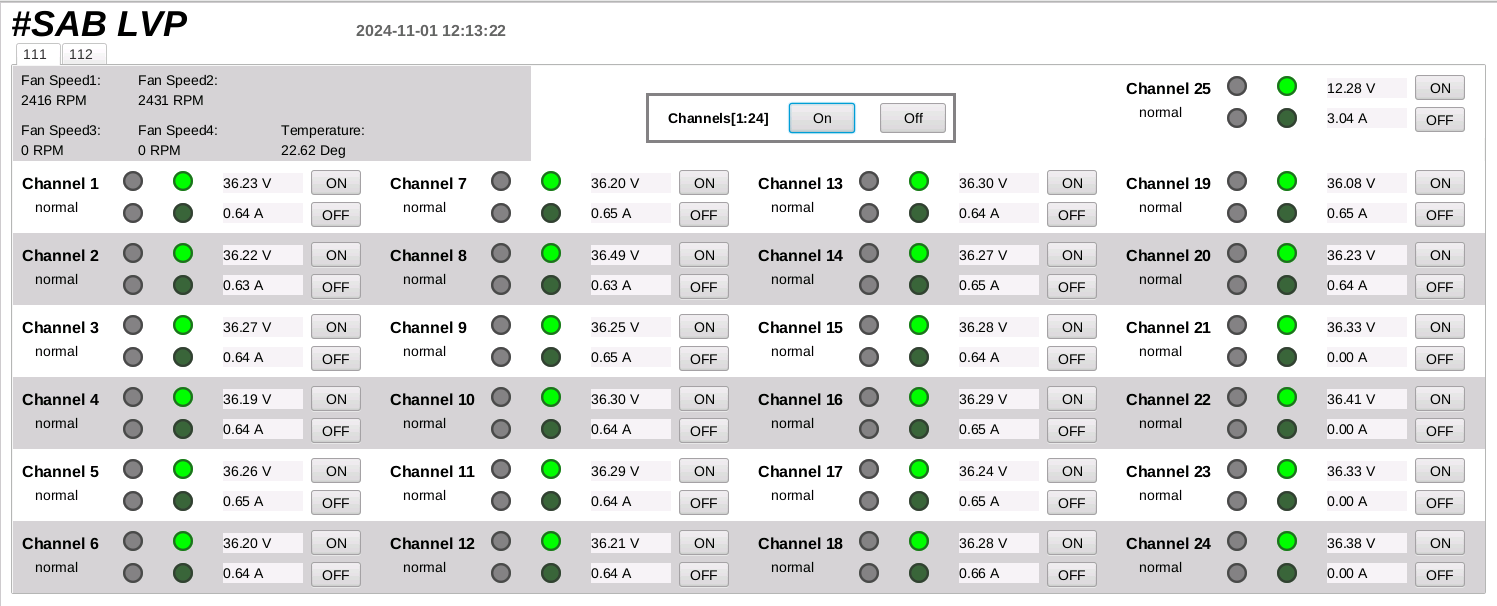}
\caption{Low voltage power vontroller for readout electronics (channels 1-24 for front-end electronics, channel 25 for BEC, with voltage and current monitored).}
\label{fig:LVP}
\end{figure}

Control system encompasses power management for the electronics and safety protocols during testing. A custom-made low-voltage power supply (50 channels, maximum output of 48 V) is utilized for the electronics and is operated via the on/off buttons of LVP software, as shown in Fig. \ref{fig:LVP}. A magnetic warning sensor is employed to control the opening and closing of the dark room doors. Monitoring for temperature and humidity in the dark room is also implemented.

\section{Test Program Design and Implementation}

\begin{figure}
\centering
\includegraphics[width=0.75\textwidth]{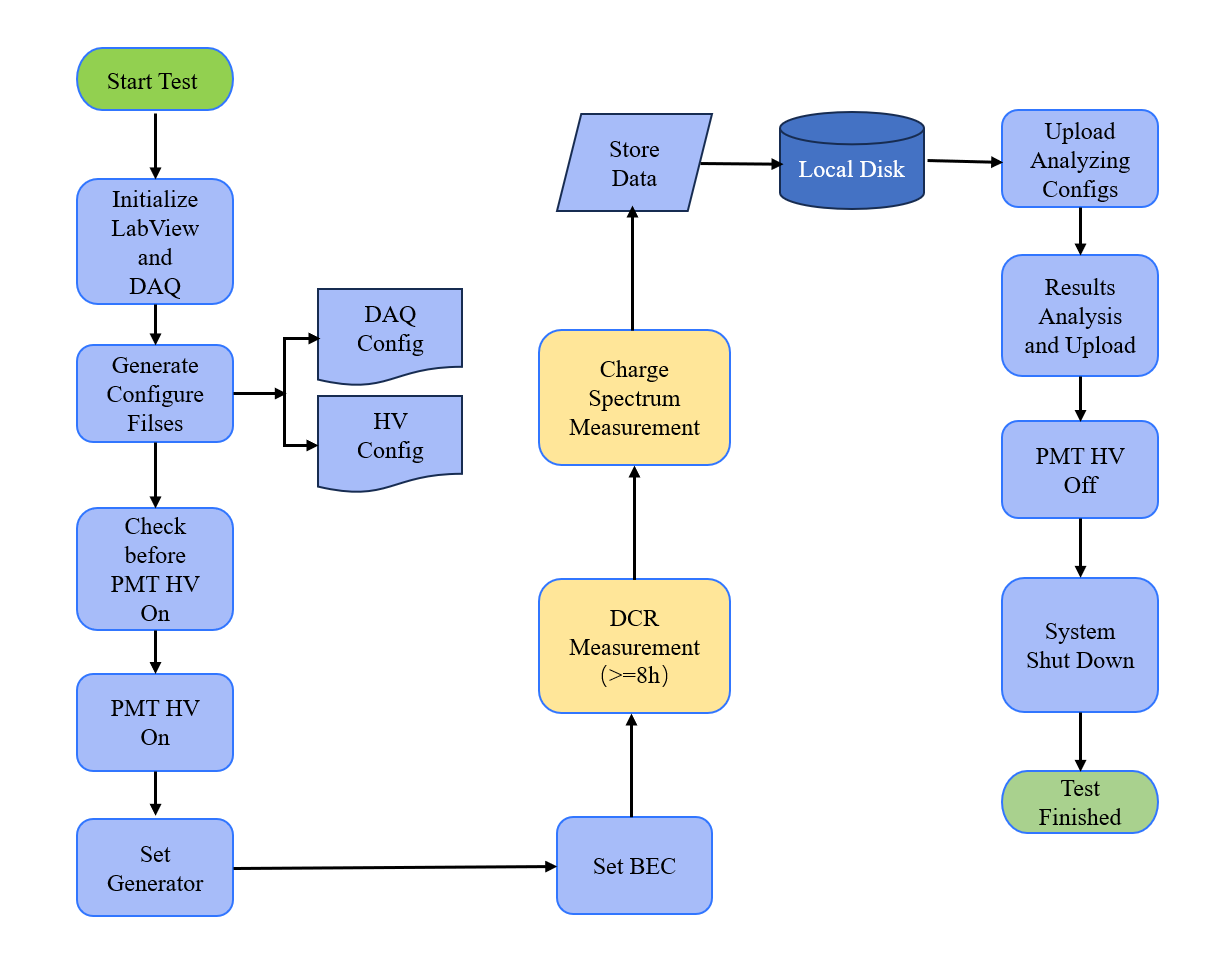}
\caption{Workflow of the test program}
\label{fig:Test_Process}
\end{figure}

The primary objectives of this test are the measurement of the DCR and the assessment of the charge spectra under LED illumination, 
as the results from these parameters are sufficient for validating PMT performance at this stage. To ensure that the total testing duration remains approximately 10 hours, allowing for two tests within a single day, the test program was designed to operate as automatically as possible. 
A detailed workflow of the test program is presented in Fig. \ref{fig:Test_Process}, note that the measurement of the DCR is under the geomagnetic field shielding and the shielding is then removed for charge spectra measurement under LED illumination. 
The most time-consuming step is the DCR measurement, which requires several hours for the PMT to cool down. For the charge spectra measurement, firstly, a baseline measurement (pedestal) is conducted with the LED turned off, followed by the measurement of charge spectra with the LED activated. 
During the test, essential information such as PMT IDs, operating voltages, and electronics channels are input via a configuration file. 
After the test is completed, the data is stored in a database on the local computer.

\begin{figure}
\centering
\includegraphics[width=0.95\textwidth]{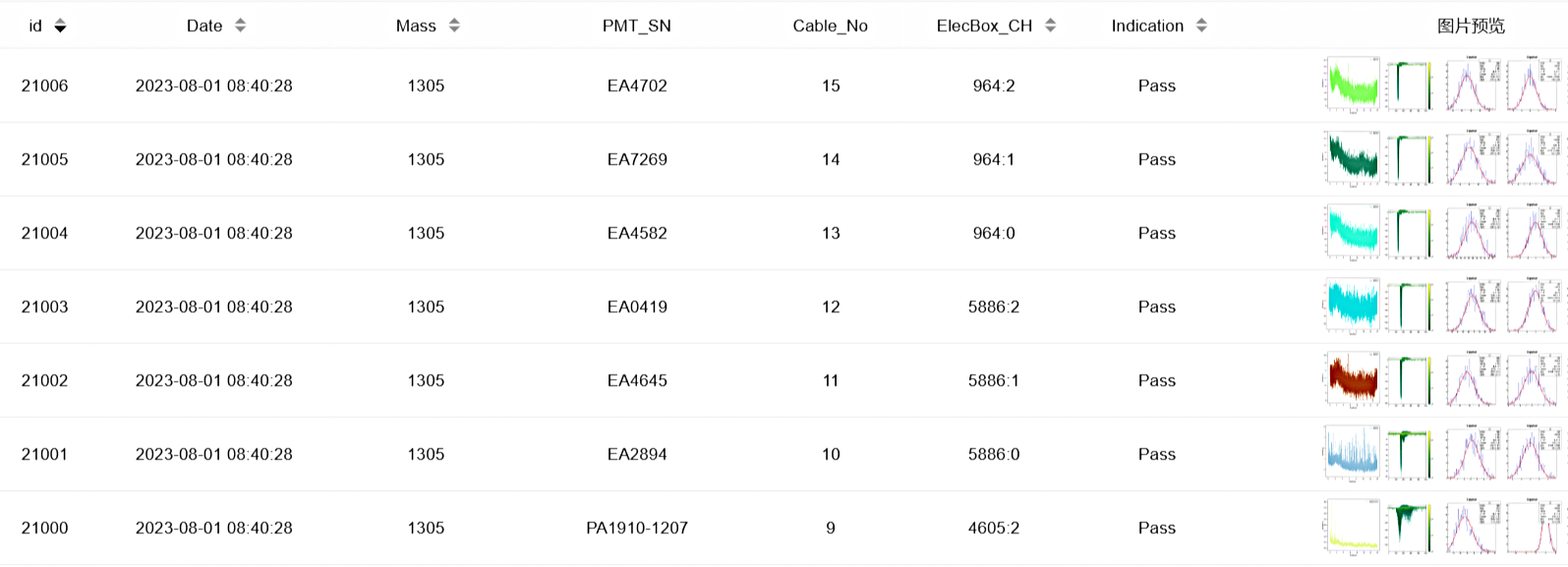}

\caption{Website displaying of test Results: includes test date, mass number, PMT serial number, number of cable connected, electronics channel, indication of test results, and plots of DCR curve and charge spectra etc..}
\label{fig:website}
\end{figure}

The implementation of the test program was conducted using LabVIEW-based process control software, which communicates with DAQ and control the test system. This software facilitates the uploading of configuration files, execution of tests, and automatic analysis of results. 
The analyzed results are uploaded to the installation database and displayed on a website, as illustrated in Fig. \ref{fig:website}. 
Moreover, the program automatically evaluates whether the PMTs meet the test criteria. Firstly, for the DCR measurement, the final DCR after cooling down must be below 100 kHz and the DCR oscillation amplitude should be less than 30\% of final DCR, with frequency less than 5 times during the test time. For the charge spectra measurement, the measured charge should be less than 0.5 pC with the LED off and larger than 3.2 pC with the LED on.

As previously indicated, the test is a part of the PMT installation process. PMTs that pass the test will be assembled into PMT modules and subsequently sent to underground for installation within a few days. Conversely, PMTs that fail the test will be inspected and retested to identify the source of the failure, whether it originates from the cable connections, the readout electronics, or the PMT itself.

\section{Performance and Initial Results of the Test System}

\subsection{DCR Cooling Down and Threshold Scanning Curve}

\begin{figure}
\captionsetup[subfigure]{labelformat=parens}
\centering
\subcaptionbox{NNVT PMT\label{fig:NNVTDCR}}{
    \includegraphics[width=0.45\linewidth]{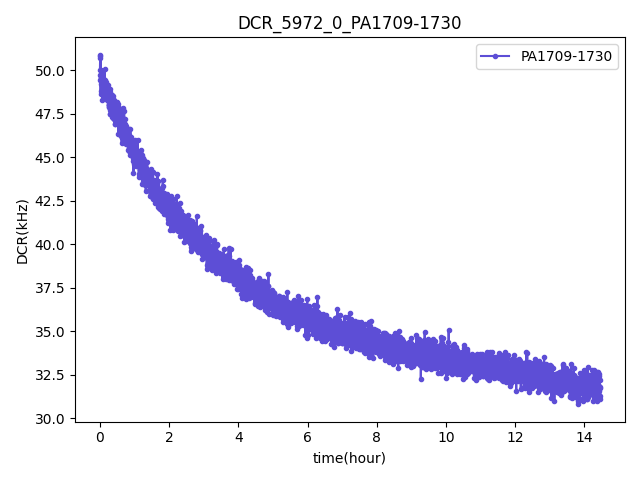}
}
\hspace{0.05\linewidth} 
\subcaptionbox{Hamamatsu PMT\label{fig:HamamatsuDCR}}{
    \includegraphics[width=0.45\linewidth]{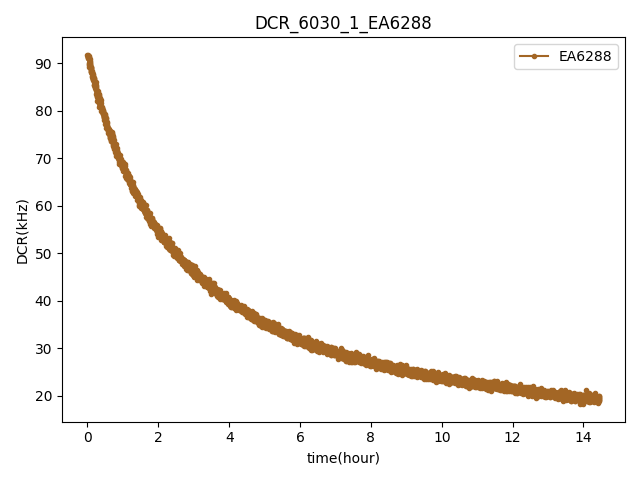}
}
\caption{Typical DCR cooling down curve over time}
\label{fig:DCRvsTime}
\end{figure}

After system setup, a small number of PMTs were initially tested to evaluate the performance of the system. 
The DCR curve as a function of time was measured for both the MCP PMT and the dynode PMT, as depicted in Fig. \ref{fig:DCRvsTime}. 
The DCR exhibits a clear downward trend as testing time increases, which is a typical characteristics: the PMTs were exposed to environmental light for several days, necessitating time for cooling before reaching a stable value. 
The duration required and the final DCR value depend on several factors, including the specific PMT, the duration of exposure, and the intensity of the incident light.

\begin{figure}
\captionsetup[subfigure]{labelformat=parens}
\centering
\subcaptionbox{NNVT PMT\label{fig:NNVTDCR_thr}}{
    \includegraphics[width=0.45\linewidth]{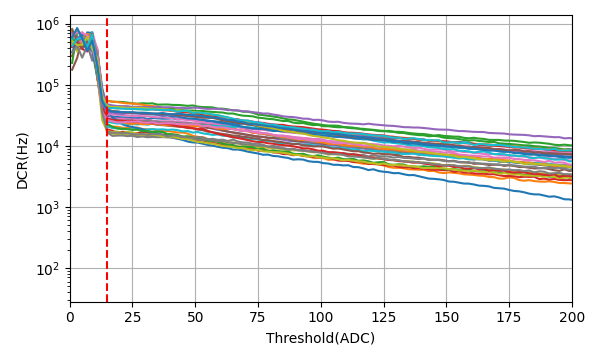}
}
\hspace{0.05\linewidth} 
\subcaptionbox{Hamamatsu PMT\label{fig:HamamatsuDCR_thr}}{
    \includegraphics[width=0.45\linewidth]{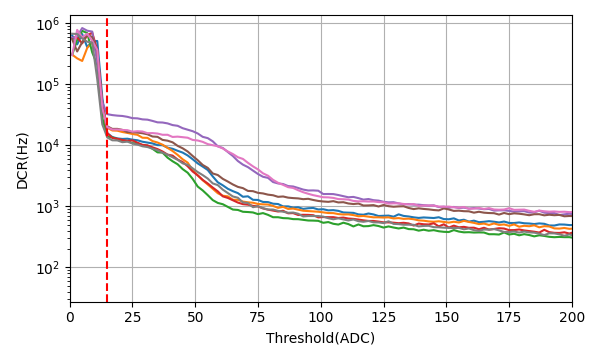}
}
\caption{DCR at different thresholds}
\label{fig:DCR_thr}
\end{figure}

To assess the overall noise level of the system, a threshold scanning of the DCR was conducted, and the results are presented in Fig. \ref{fig:DCR_thr}. 
Different types of PMTs exhibit slightly different curves; 
however, a rapid change in DCR occurs around 15 ADCs (corresponding to 0.25 photoelectrons (p.e.) at gain 1E7) for both PMTs. 
This value has been established as the threshold for subsequent DCR measurements. 
Additionally, the dynode PMT shows a notable change between 50 to 60 ADCs ($\approx 1 \text{ p.e.}$), which is less pronounced in the MCP PMT, due to the known long tail in its charge distribution \cite{17,18}.

\subsection{DCR with the Varied Geomagnetic Field Shielding}
\begin{figure}
\captionsetup[subfigure]{labelformat=parens}
\centering
\subcaptionbox{NNVT PMT\label{fig:DCR as a function of EMF for MCP PMT}}{
\includegraphics[width=0.45\textwidth]{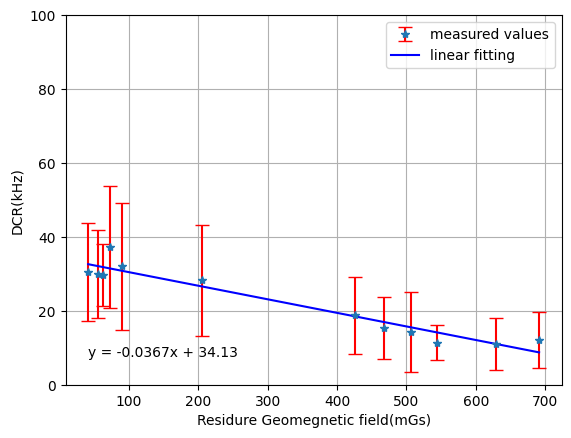}
}
\hspace{0.05\linewidth} 
\subcaptionbox{Hamamatsu PMT\label{fig:DCR as a function of EMF for dynode PMT}}{
\includegraphics[width=0.45\textwidth]{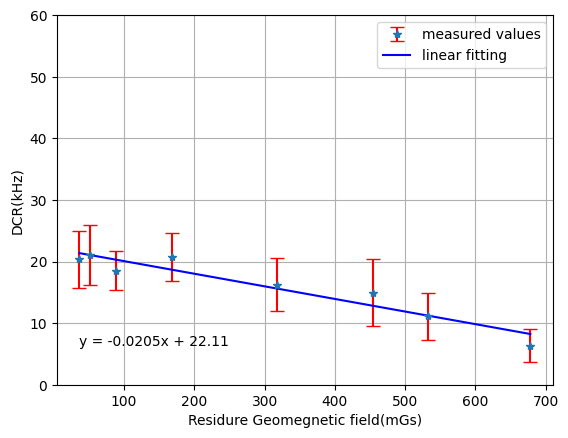}
}
\caption{DCR as a function of the geomagnetic field}
\label{fig:DCRvsEMF}
\end{figure}

As mentioned in Section 2.2, various shielding methods were explored to mitigate the effects of the geomagnetic field, during which a number of PMTs were evaluated. Additionally, the magnetic field in the darkroom were found to be somewhat non-uniform, as demonstrated in Fig. \ref{fig: residual EMF}. Consequently, DCR were recorded under different magnetic field. It was observed that DCR is highly sensitive to the magnetic field. This is illustrated in Fig. \ref{fig:DCR as a function of EMF for MCP PMT} and Fig. \ref{fig:DCR as a function of EMF for dynode PMT}, which depict the relationship for both MCP and dynode PMTs, respectively. Linear fits are included in both plots, which can be used for correction of the DCR meausred at different magnetic field. The data indicate that DCR decreases as the magnetic field strength increases. This feature can be attributed to the fact that the DCR in PMTs primarily arises from thermal electron emission from the photocathode, with collection efficiency decreasing in stronger magnetic fields. Furthermore, the impact of the magnetic field on MCP PMTs was found to be slightly more pronounced than that on dynode PMTs. 
This study provides a useful reference for DCR measurement in JUNO acutual setup.

\subsection{DCR Assessment of Initial PMT Samples}
\begin{figure}
\captionsetup[subfigure]{labelformat=parens}
\centering
\subcaptionbox{DCR distribution of initial PMT samples\label{fig:Summary of DCR}}{
\includegraphics[width=0.45\textwidth]{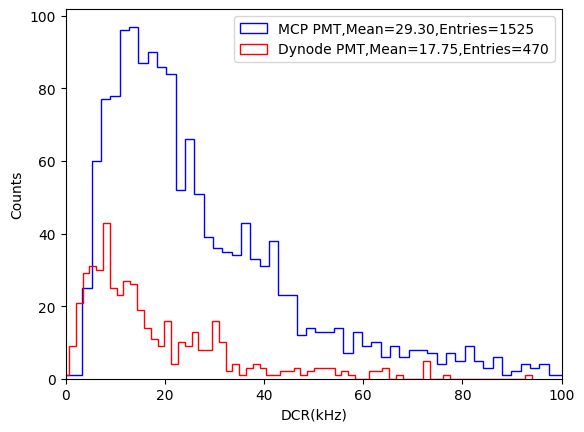}
}
\hspace{0.05\linewidth} 
\subcaptionbox{DCR differs in test systems\label{fig:test systems differences}}{
\includegraphics[width=0.45\textwidth]{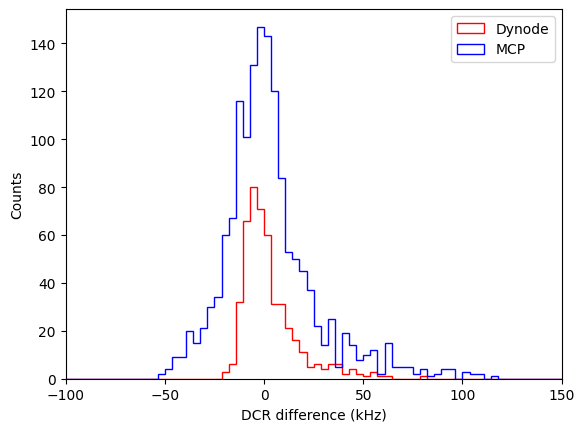}
}
\caption{DCR test results of initial PMT samples}
\label{fig:Summary of DCR in SAB}
\end{figure}

The DCR distribution for the initial evaluation of about 2000 PMTs is presented in Fig. \ref{fig:Summary of DCR}, which includes data for both MCP and dynode PMTs, after correcting for the effects of the geomagnetic field for those PMTs without shielding (several hundred PMTs at the beginning of the test ), using the fitted paramers in Fig. \ref{fig:DCRvsEMF}. 
The averaged values of distributions are summarized in Tab. \ref{tab:DCR_Initial}.
For comparative purposes, test results from Pan-Asia are included in the table \cite{14}. 
Additionally, a plot illustrating the DCR differences of individual PMTs from the two tests is provided in Fig. \ref{fig:test systems differences} for closer check. 
The data presented in the table and figures indicates a general agreement between the two tests, with no significant systematic shifts observed. However, a small number of PMTs exhibit relatively large discrepancies, which can be attributed to variations in test conditions such as cooling down time, temperature, and other factors.
This consistency suggests that the performance of the PMTs is not changed upon arrival at the JUNO. 
Only two of the initial 2000 PMTs were found to have very high and unstable DCR ($> 1$ MHz), which need further study and were decided not to use them for installation.

\begin{table}[htp]
\begin{center} 
\caption{DCR (kHz) average values of initial PMT samples}
\label{tab:DCR_Initial}
\begin{tabular}{c|cc} 
PMT Type & Pan Aisa & JUNO SAB \\ 
\hline 
NNVT & 32.4 & 29.3 \\ 
Hamamatsu & 16.6 & 17.7 \\ 
\end{tabular}
\end{center}
\end{table}

\subsection{Charge Spectra Measurement of Initial PMT Samples}
\begin{figure}
\captionsetup[subfigure]{labelformat=parens}
\centering
\subcaptionbox{A typical signal of PMT\label{fig:a typical signal of PMT}}{
\includegraphics[width=0.45\textwidth]{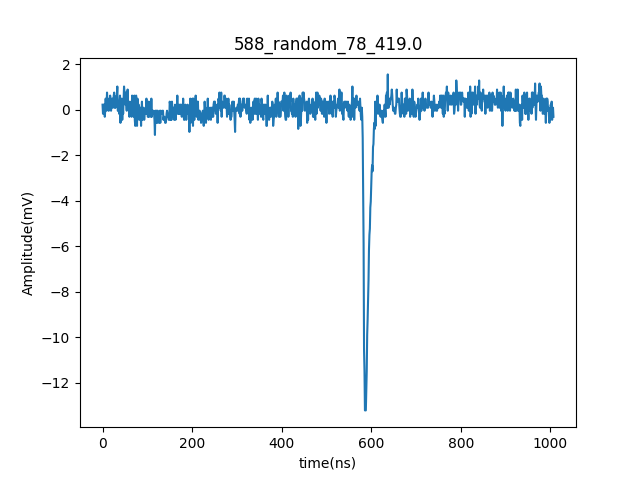}
}
\hspace{0.05\linewidth} 
\subcaptionbox{The charge spectra of PMT\label{fig:charge_spectrum}}{
\includegraphics[width=0.45\textwidth]{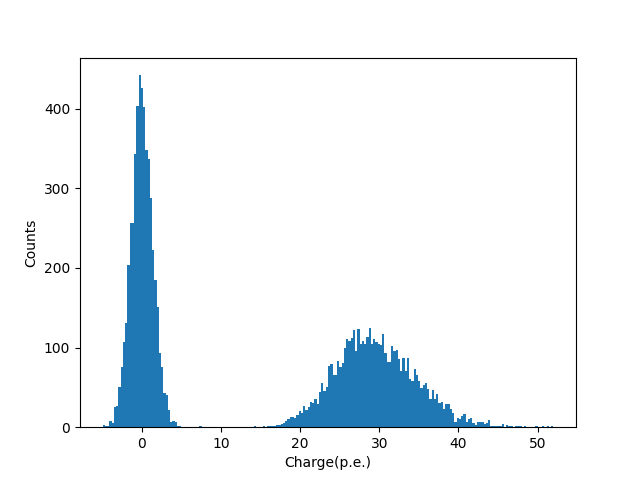}
}
\caption{Signal and charge spectra}
\label{fig:signal_chargespectrum}
\end{figure}

To evaluate the system's capability for charge spectra measurement, PMTs were tested under LED illumination. 
It need to know that the test was conducted in the presence of the geomagnetic field without the shielding box, as the LEDs were designed to be positioned at the top of the dark room.
A representative waveform from PMT, with an amplitude of approximately 12 mV, is shown in Fig. \ref{fig:a typical signal of PMT}. 
The charge spectra distribution is presented in Fig. \ref{fig:charge_spectrum}, with the first peak corresponding to the pedestal and the second peak representing the measured charge from LEDs, which is approximately 30 p.e. 

\begin{figure}
\centering
\includegraphics[width=0.75\textwidth]{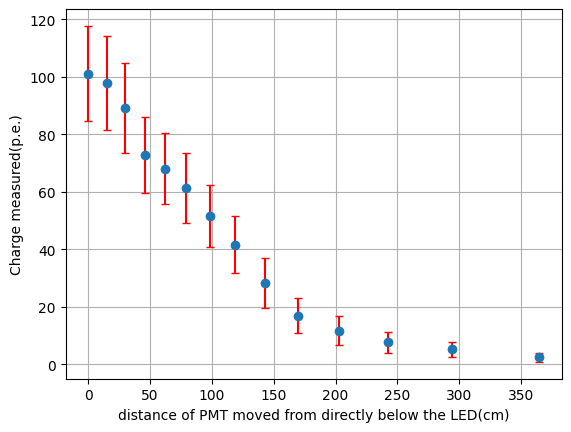}
\caption{Charge variation as a function of a horizontal distance of PMT}
\label{fig:charge_distance}
\end{figure}

As previously mentioned, LEDs were employed for illumination to assess the PMT's response to light. Due to varying distances between the different PMTs and the LED, the collected charge at each distance was measured, as depicted in Fig. \ref{fig:charge_distance}, where the maximum of measured charge is 100 p.e., when the PMT is placed directly below the LED (about 170 cm below), and the charge reduces to single photon level when PMT is moved to a horizontal distance of 350 cm away.
 
\begin{figure}
\centering
\includegraphics[width=0.75\textwidth]{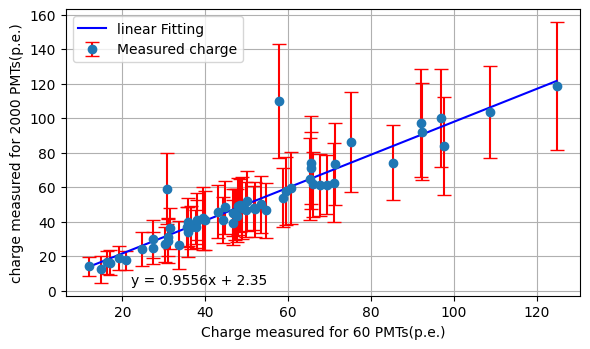}
\caption{Measured charge at different test channels for the first 60 and 2000 PMTs}
\label{fig:MCP_PE_fit}
\end{figure}
The consistency of light response of different PMTs is also checked, as depicted in Fig. \ref{fig:MCP_PE_fit}, where the x-axis indicates the charge measured for 60 PMTs at different horizontal distance at the first test and y-axis represents the charge measured for 2000 PMTs at the later tests.
A linear fitting is applied and its slope is about 1, representing the PMTs have the consistent response for charge measurement.

\section{Summary and Conclusion}
A test system was developed at the JUNO site prior to PMT installation, with initial samples of the PMTs being tested. This paper details the design and implementation of the system, which includes a dark room with the geomagnetic field shielding, readout electronics, data acquisition and control, test program and procedures. The dark room was meticulously designed to prevent light leakage and minimize the impact of the geomagnetic field. The JUNO readout electronics and a prototype of the JUNO DAQ were utilized, and the test program and procedures were optimized for PMT installation.

Upon completion of the test system, the system's performance was evaluated using initial PMT samples. 
This evaluation included measurements of the DCR cooling down and threshold scanning curves, DCR under different magnetic fields, assessment of DCR with initial 2000 PMTs samples, and PMTs charge spectra under LED illumination. 
The test results indicated that the system is functioning effectively, and the functionality of the PMTs is good and consistent to that from the Pan-Asia test. 
Only a few PMTs ($\approx 0.1\%$) are observed with abnormal DCR and under study. More test results will come in future paper. Installation of the JUNO PMTs is currently underway, and this system plays a crucial role in verifying the functionality of the PMTs prior to their installation.


\printbibliography[heading=bibintoc, title=\ebibname]

\appendix

\addappheadtotoc

\end{document}